\documentclass[epj,twocolumn]{webofc} 
\usepackage[varg]{txfonts}   
\usepackage{hyperref}
\usepackage{textcomp}
\usepackage{color}

\woctitle{ISVHECRI 2018}

\begin{document}
\title{Air Shower Detection by Arrays of Radio Antennas}

\author{\firstname{Frank G.} \lastname{Schr\"oder}\inst{1,2}\fnsep\thanks{\email{fgs@udel.edu} ~~ and ~~\email{frank.schroeder@kit.edu}}
}

\institute{Bartol Research Institute, Department of Physics and Astronomy, University of Delaware, Newark, DE, USA
           \and
           Institute for Nuclear Physics, Karlsruhe Institute of Technology (KIT), Karlsruhe, Germany  
          }

\abstract{
Antenna arrays are beginning to make important contributions to high energy astroparticle physics supported by recent progress in the radio technique for air showers. 
This article provides an update to my more extensive review published in Prog.~Part.~Nucl.~Phys.~93 (2017) 1. 
It focuses on current and planned radio arrays for atmospheric particle cascades, and briefly references to a number of evolving prototype experiments in other media, such as ice. 
While becoming a standard technique for cosmic-ray nuclei today, in future radio detection may drive the field for all type of primary messengers at PeV and EeV energies, including photons and neutrinos. 
In cosmic-ray physics accuracy becomes increasingly important in addition to high statistics. 
Various antenna arrays have demonstrated that they can compete in accuracy for the arrival direction, energy and position of the shower maximum with traditional techniques. 
The combination of antennas and particles detectors in one array is a straight forward way to push the total accuracy for high-energy cosmic rays for low additional cost.
In particular the combination of radio and muon detectors will not only enhance the accuracy for the cosmic-ray mass composition, but also increase the gamma-hadron separation and facilitate the search for PeV and EeV photons.
Finally, the radio technique can be scaled to large areas providing the huge apertures needed for ultra-high-energy neutrino astronomy.
}
\maketitle
\section{Introduction}
Building on the progress in the digital techniques for radio detection of air showers, antenna arrays are beginning to make important contribution to high-energy astroparticle physics.
As outlined in two recent review article \cite{SchroederReview2016, HuegeReview2016}, several digital radio arrays successfully measure cosmic-ray air showers since the last decade. 
Direct comparisons of the measurements of radio arrays and traditional air-shower detectors have experimentally shown that radio arrays can deliver a competitive precision for the arrival direction, energy and position of the shower maximum.
The collaboration of experimentalists and model builders lead to a detailed understanding of the mechanisms of the radio emission by air showers. 
Monte Carlo simulation codes of today, such as CoREAS \cite{HuegeCoREAS_ARENA2012} and ZHAires \cite{Alvarez_ZHAires_2012}, have been tested to describe measured radio amplitudes within the experimental uncertainties of $10-20\,\%$ \cite{2015ApelLOPES_improvedCalibration, TunkaRex_NIM_2015, AERAenergyPRL2015}.
In turn, simulations are now used to interpret measurements by radio arrays, and to reconstruct the atmospheric depth of the shower maximum, $X_\mathrm{max}$, with high precision.

\begin{figure*}
  \centering
  \includegraphics{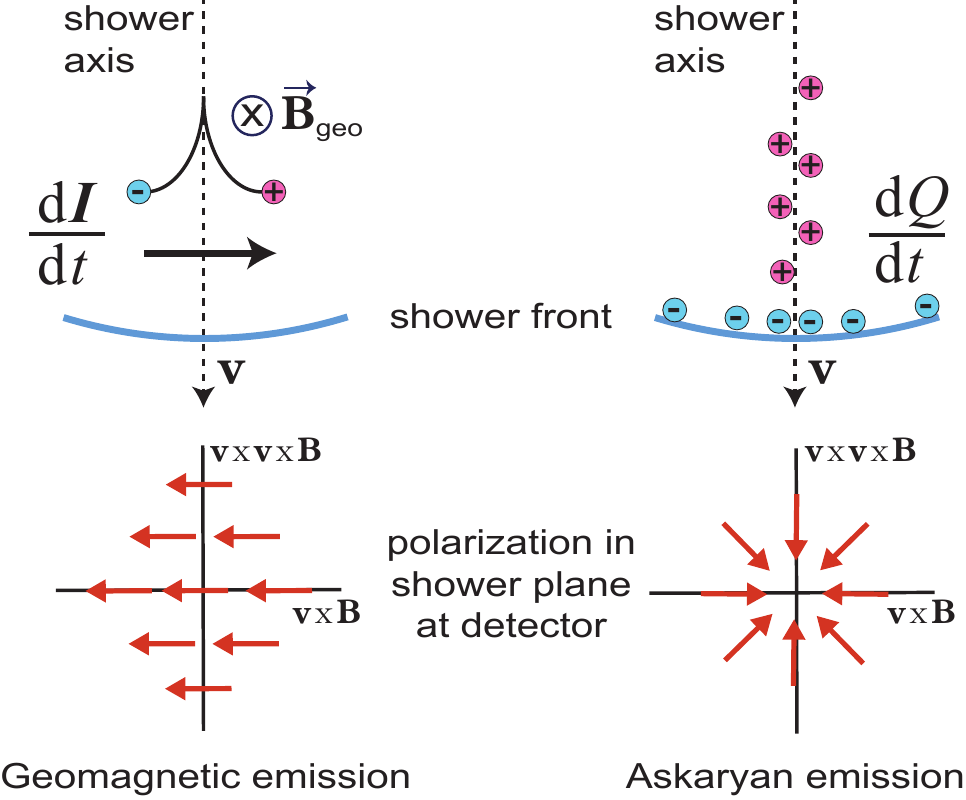}
  \caption{Two mechanisms dominate the radio emission of air showers. 
  The deflection of electrons and positrons by the geomagnetic Lorentz force induces a time-varying transverse current causing radio emission in the $v \times B$ direction. 
  By ionizing the traversed air, the shower front builds up a time-varying net charge excess causing radially polarized radio emission, also called Askaryan emission. 
  Not shown: In thunderstorm clouds particle acceleration by electric fields is a third mechanism \cite{2011ApelLOPES_Thunderstorm, SchellartLOFARthunderstorm2015} (adapted from \cite{SchroederReview2016}).}
  \label{fig_emission}
\end{figure*}

\section{Basic properties of the radio signal}
At least two mechanisms are relevant for the radio emission by air showers: 
geomagnetic and Askaryan emission (Fig.~\ref{fig_emission}). 
The generally stronger geomagnetic emission is caused by the time variation of a transverse current induced by the geomagnetic field in the shower front. 
Askaryan emission is caused by a negative charge excess in the shower front. 
Both the net charge excess and its time-variation lead to coherent radiation, with the latter being the dominant effect in the atmosphere at $100\,$MHz \cite{JamesPC2018}.
Askaryan and geomagnetic emission are both linearly polarized, the geomagnetic one in the $v \times B$ direction ($v$ is the direction of the shower axis and $B$ the geomagnetic field), and the Askaryan emission radially with respect to the shower axis \cite{AugerAERApolarization2014, SchellartLOFARrization2014}. 
The maximum emission is close to the shower maximum, and the relative emission strength of both mechanisms seems to vary during the shower development \cite{Glaser_ShowerModel_2016}. 
Consequently, not only the relative strengths of both mechanism varies with distance to the shower axis, but also the resulting polarization does.
Since the resulting polarization is the sum of two linear polarization, it was assumed to be approximately linear, too, until a few years ago. 
However, the total signal has been measured to feature a small circular polarization comportment \cite{LOFAR_circularPolarization_PRD2016} explained by a small phase shift between the geomagnetic and Askaryan emission \cite{Glaser_ShowerModel_2016}.
Except for selected geometries, the Askaryan effect generally is weaker than the geomagnetic one. 
This means that for first-order approximations, the Askaryan effect often can be ignored, but it has to be taken into account for an accurate reconstruction of air-shower parameters.

Apart from the polarization and the implied interference of geomagnetic and Askaryan emission, the characteristics of the radio signal on ground do not depend too much on emission mechanism. 
Due to relativistic beaming the radio signal is emitted in a forward cone of $2-3^\circ$ opening angle for all emission mechanisms. 
For this reason, the size of the footprint depends drastically on the distance to the shower maximum and, thus, mainly on the zenith angle and to a smaller extent on the altitude of the detector.
The lateral extension of the air-shower usually can be neglected, except for detailed structures. 
Most of the electrons and positrons contributing to the radio emission are within a few $10\,$m distance from the shower axis (not few m as wrongly written in my review \cite{SchroederReview2016}). 
Although this lateral extension of the emission region is not negligible against the wavelengths it is small compared to the typical dimension of radio arrays ranging from $100\,$m to many km.
Finally, the size of the radio footprint is almost independent of the energy of the primary particle. 
The radio footprint has a diameter only $200-300\,$m for vertical showers and many kilometers for near horizontal showers \cite{AERAinclined_JCAP2018}.

The shape of the radio footprint is asymmetric along $v \times B$ direction because the interference between geomagnetic and Askaryan emission is constructive in one direction and destructive in the opposite direction.
Depending on the distance to the shower maximum \cite{AllanICRC1971, Glaser_LDF_2018}, the radio amplitude either falls approximately exponentially with increasing distance to the shower axis, or experiences a ring-like enhancement at the Cherenkov-angle and drops exponentially further out. 

Finally, not only the amplitude and polarization structure has been measured, but also the wavefront of the radio signal could be measured thanks to nanosecond precise time calibration of selected antenna arrays \cite{2014ApelLOPES_wavefront, CorstanjeLOFAR_wavefront2014}. 
According to these LOPES and LOFAR measurements of individual events, the shape of the radio wavefront can better be approximated by a hyperboloid than by a plane, cone, or a sphere. 
It has not yet been understood, why CODALEMA measurements of the average radio wavefront could not confirm these results \cite{CODALEMAwavefront_ICRC2017}. 
In any case, an accurate description of the wavefront is mandatory only when aiming at an accuracy for the arrival direction of better than $1^\circ$.

\begin{figure}
  \centering
  \includegraphics[width=\linewidth]{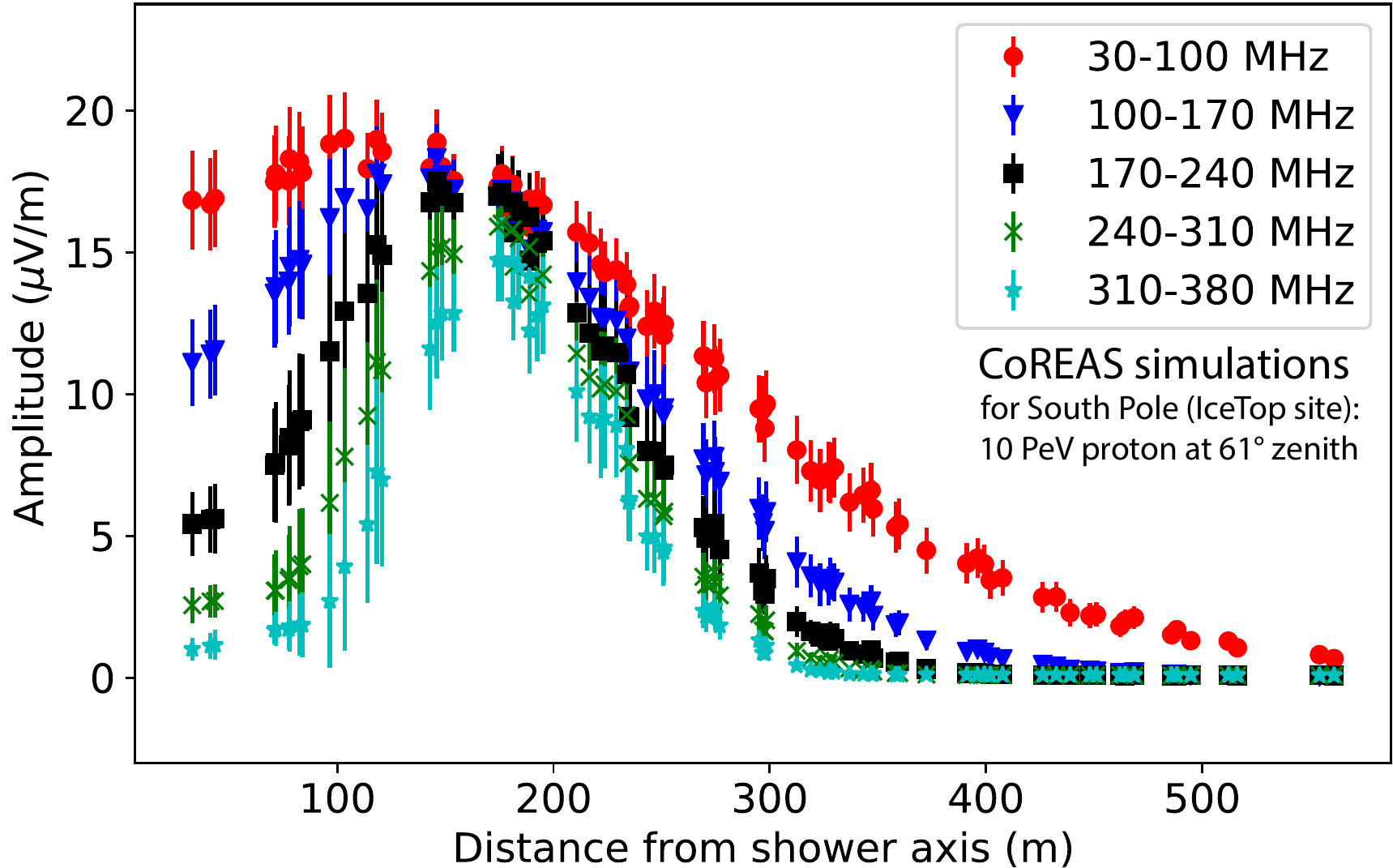}
  \caption{Lateral distribution of the amplitude of the radio signal in different frequency band. 
The bump corresponding to a ring at the Cherenkov angle. 
Here the radio signal is nearly equally strong in all frequency bands. 
Only at frequencies $\lesssim 200\,$MHz the ring is filled, i.e., detection at higher frequencies requires antennas directly at the Cherenkov ring (adapted from \cite{Balagopal2018})}.
  \label{fig_latFreqPlot}
\end{figure}

The frequency band of the radio emission earlier was thought to be mainly below $100\,$MHz, because the wavelength then is larger than the thickness of the shower front.
However, since the radio signal propagates approximately with the same speed as the shower front (both slightly slower than the speed of light in vacuum), the emission can be coherent even at higher frequencies (Fig.~\ref{fig_latFreqPlot}). 
In particular at the Cherenkov angle, the emission extents to several GHz \cite{CROME_PRL2014}. 
Since in radio quiet areas the dominant background is Galactic noise dropping rapidly with increasing frequency, this insight led to a re-investigation of the optimum frequency band.
While many previous and current arrays have frequency bands around $30-80\,$MHz, it now seems that measuring between $100-200\,$MHz will provide a significantly better signal-to-noise ratio \cite{Balagopal2018}.

\section{State of the art experiments}
In the last fifteen years a number of dedicated antenna arrays was build for the detection of air showers. 
These consisted of either extending existing air-shower arrays by radio antennas or by enhancing astronomical antenna arrays so they can be used for air-showers detection in parallel to astronomical observations. 
Only recently the community has started to build stand-alone antenna arrays successfully measuring air showers \cite{ARIANNA_2016}. 
Thanks to the progress made by hybrid arrays of radio and particle detectors, this now is possible.
It makes the radio technique an economic alternative for cosmic-ray detection with high statistics. 
Nevertheless, for science goals requiring the best achievable accuracy, the combination of particle and radio detectors remains the method of choice. 

Essential pathfinders for the digital radio technique were LOPES, the radio extension of the KASCADE-Grande detector \cite{FalckeNature2005, Apel2010KASCADEGrande}, and CODALEMA, a dedicated radio array in France \cite{ArdouinBelletoileCharrier2005}. 
CODALEMA contributed significantly to the current understanding of the emission mechanisms \cite{Ardouin2009, CODALEMAchargeExcess2015}. 
Despite in a radio-loud environment, LOPES provided several proof-of-principle demonstrations, such as experimental evidence for the sensitivity of radio measurements to the longitudinal shower development \cite{2012ApelLOPES_MTD, 2014ApelLOPES_MassComposition}.
Furthermore, LOPES pioneered interferometric radio detection of air showers by cross-correlation beamforming, that was possible thanks to a nanosecond precise time-calibration using a beacon \cite{SchroederTimeCalibration2010}. 
While the benefit of interferometric techniques is not obvious for sparse arrays with spacings larger than $100\,$m, it can lower the threshold when at least a few antennas are located close to each other \cite{ARA_interferometricTrigger_2018}. 
Even though not for interferometry, the beacon method is still used at the sparser Auger Engineering Radio Array (AERA) for time calibration \cite{AERAairplanePaper2015}.

LOFAR, Tunka-Rex, and AERA are second-generation radio arrays built on the experience of LOPES and CODALEMA. 
They could demonstrate that the radio technique can compete in measurement accuracy with established techniques, and now is mature for its application on cosmic-ray science. 

\begin{figure}
  \centering
  \includegraphics[width=\linewidth]{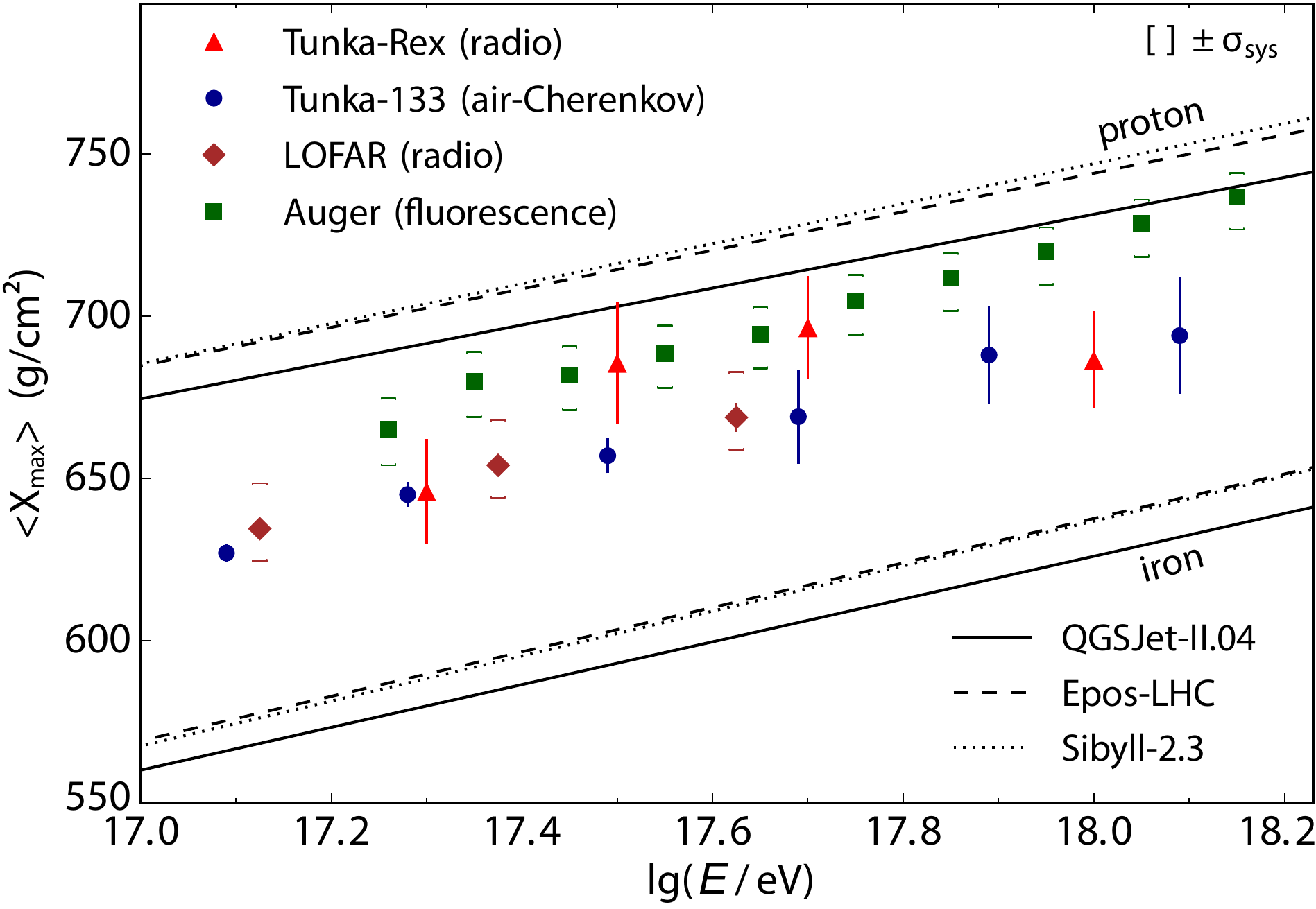}
  \caption{Mean $X_\mathrm{max}$ over energy measured by three different techniques: fluorescence light (Auger \cite{AugerXmax_ICRC2017}), air-Cherenkov light (Tunka-133 \cite{Tunka133_TAUP2015}), and radio (LOFAR \cite{LOFARNature2016}, Tunka-Rex \cite{TunkaRexPRD2018}). 
  Systematic uncertainties have been studied extensively for the fluorescence technique since more than a decade, and for the radio technique since several years by LOFAR (adapted from \cite{TunkaRexPRD2018})}.
  \label{fig_xmaxOverview}
\end{figure}

LOFAR is a European radio observatory whose dense core in the Netherlands is triggered by a dedicated particle-detector array for air-shower measurements \cite{SchellartLOFAR2013}.
Due to the dense antenna spacing LOFAR provides the most detailed measurements of the radio signal of air showers, e.g., a measurement of the relative strengths of Askaryan and geomagnetic emission as function of zenith angle and distance to the shower axis \cite{SchellartLOFARrization2014}. 
Moreover, LOFAR features the most precise measurements the atmospheric depth of the shower maximum, $X_\mathrm{max}$, reconstructed by comparing measured radio signals to CoREAS simulations \cite{BuitinkLOFAR_Xmax2014} (Fig.~\ref{fig_xmaxOverview}).

Tunka-Rex is the radio extension of the Tunka-133 non-imaging air-Cherenkov array in Siberia \cite{TunkaRex_NIM_2015}, and meanwhile is also triggered by the co-located particle-detector array Tunka-Grande \cite{TAIGA_2014}. 
By direct comparison with the air-Cherenkov measurements, Tunka-Rex could experimentally cross-check the energy and $X_\mathrm{max}$ reconstruction and their accuracies \cite{TunkaRex_Xmax2016}. 
When exploiting the information of the pulse shape in each antenna, the $X_\mathrm{max}$ accuracy is roughly equal to the air-Cherenkov technique and the energy precision is even better, limited mostly by systematic uncertainties of the antenna gain \cite{TunkaRexPRD2018}. 
Finally, Tunka-Rex together with LOPES demonstrated that radio extensions provide an excellent tool for comparing the absolute energy scales of different air-shower arrays \cite{TunkaRexScale2016}.

\begin{figure*}[t]
  \centering
  \includegraphics[width=\linewidth]{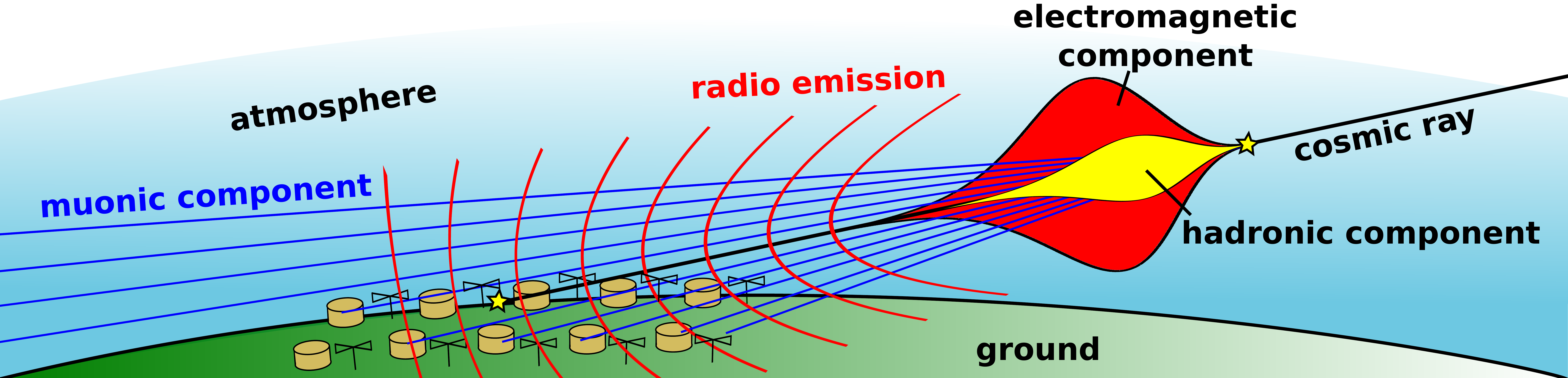}
  \caption{Sketch of an inclined air shower. 
  Only muons and the radio emission from the electromagnetic shower component survive until ground. 
  Due to the large size of the radio footprint of inclined showers, the antenna spacing can be of $o(1\,$km$)$, i.e., radio antennas can be placed with the same spacing as particle detectors (from \cite{HoltThesis2018})}.
  \label{fig_inclinedShowerSketch}
\end{figure*}

With more than 150 antennas of different types \cite{AERAantennaPaper2012} on $17\,$km$^2$ AERA currently is the largest antenna array for air showers \cite{Holt_AERA_ICRC2017}. 
AERA consists of autonomous stations and is one of the low-energy enhancements of the Pierre Auger Observatory in Argentina \cite{AugerNIM2015}. 
Although featuring a self-trigger in addition to external triggering by the Auger surface and fluorescence detectors, most of the scientific results were acquired using the trigger by the surface detectors. 
AERA confirmed the sensitivity of the radio signal to $X_\mathrm{max}$ using coincident radio and fluorescence measurements \cite{Holt_AERA_ICRC2017}.
Thanks to an elaborated antenna calibration \cite{AERAantennaCalibration2017}, AERA became also one of the most accurate detectors for the cosmic-ray energy by exploiting that the total radiation energy is proportional to the energy content of the electromagnetic shower component \cite{AugerAERAenergy2015}. 

In summary, LOFAR, Tunka-Rex, and AERA have experimentally shown that the most important shower quantities can be measured with sufficient accuracy: the arrival direction to better than a degree, the energy with a precision and accuracy of better than $15\,\%$, and $X_\mathrm{max}$ with an accuracy of $20-30\,$g/cm$^2$. 
Thus, the demonstrated resolution of radio detection is similar to the one of the established air-Cherenkov and fluorescence techniques.
Consequently, above its detection threshold the radio detection now provides a 24/7 alternative not restricted to clear nights.

\section{Planned Radio Arrays for Air Showers}
This section gives an overview of selected planned and proposed antenna arrays for air-shower detection. 

\subsection{Auger Radio Upgrade}
The Pierre Auger Observatory currently undergoes a major upgrade, called AugerPrime, improving all of it detector systems \cite{AugerPrime2015}. 
The $3000\,$km$^2$ large surface array of water-Cherenkov detectors will be upgraded by additional scintillation detectors to better distinguish the electromagnetic from the muonic shower component. 
Additionally the enhancement area, where the current AERA is located, will be equipped with underground muon detectors \cite{AugerAMIGAprototype2016}.

\begin{figure}
  \centering
  \includegraphics[width=\linewidth]{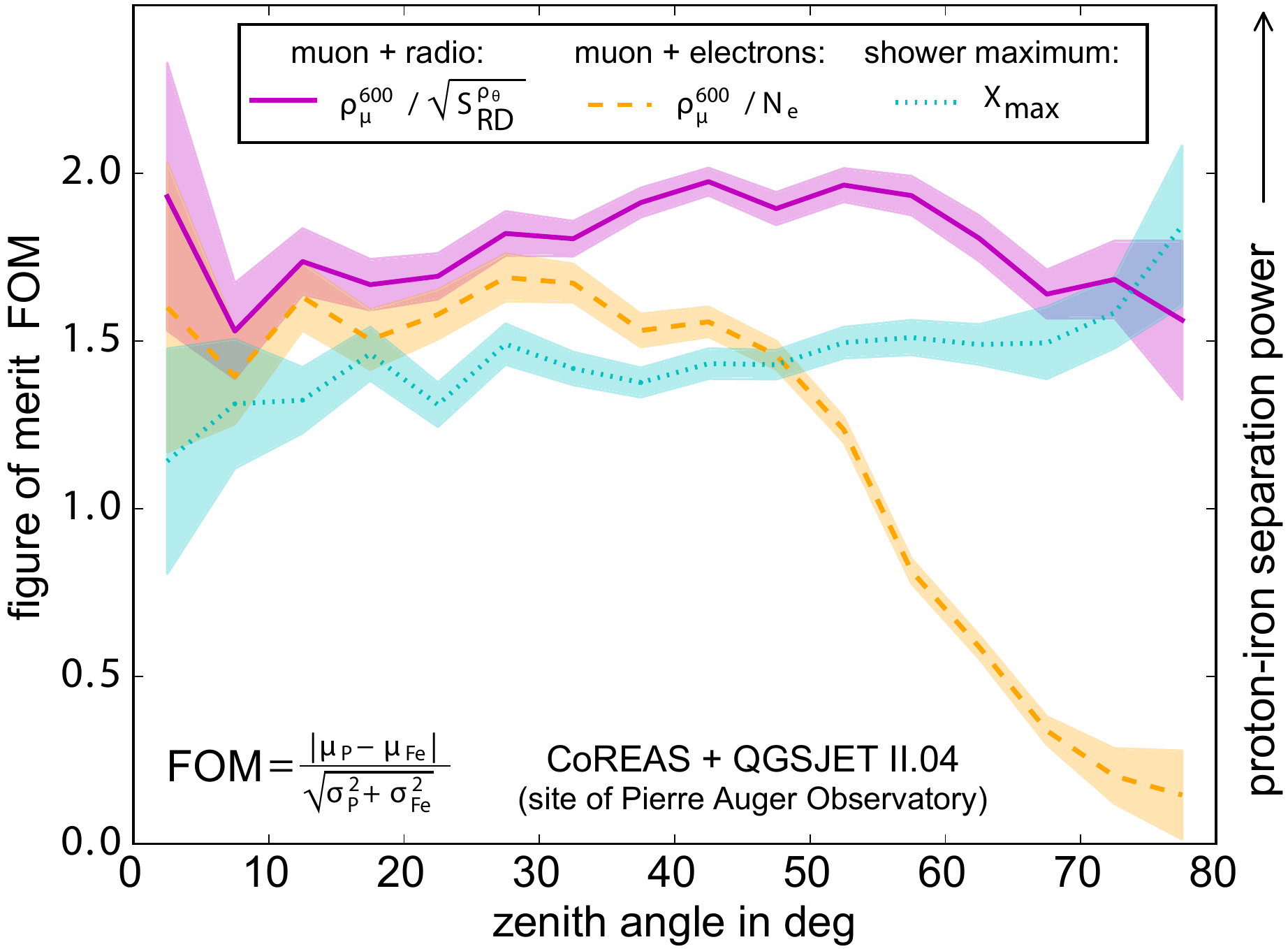}
  \caption{Figure of merit as a measure for the performance of the separation of protons from iron nuclei for different methods. 
  While the traditional combination of electron and muon detectors is a good choice for less inclined showers, the combination of radio and muon detectors is sensitive to the mass composition for all zenith angles.
  The measurement of the shower maximum provides additional information on the mass composition for all zenith angles. 
  This figure includes intrinsic uncertainties due to shower-to-shower fluctuations, but no detector-related uncertainties such as background (adapted from \cite{Holt_ARENA2018}).}
  \label{fig_RDMD_FOMoverZenith}
\end{figure}

A recent simulation study confirmed that the combination of radio and muon measurements provides a complementary method for the measurement of the  mass composition that can be combined with $X_\mathrm{max}$. 
This is comparable to the traditional method of electron and muon measurements that, however, does not work for inclined showers because the electrons are absorbed in the atmosphere.
For inclined showers only muons and the radio emission reaching ground and can be detected (see Fig.~\ref{fig_inclinedShowerSketch}).
Thus, a radio extension of muon detectors is expected to provide mass-sensitivity even for near-horizontal showers (see Fig.~\ref{fig_RDMD_FOMoverZenith}).

This is why the AugerPrime upgrade now foresees also radio antennas: each water-Cherenkov tank will be supplemented by a radio antenna in addition to the upgrade by a scintillation detector \cite{AugerRadioUpgrade_ARENA2018}.
The threshold in energy and zenith angle for the radio upgrade have yet to be determined precisely. 
Nevertheless, measurements of inclined showers by AERA have confirmed the expected large radio footprints enabling their detection by a spare antenna array. 
For zenith angles larger than $60^\circ$ the average diameter of the radio footprint exceeds $1\,$km in the direction perpendicular to the shower axis \cite{AERAinclined_JCAP2018}, and is even larger in other directions due to projection effects. 

\begin{figure*}[t]
  \centering
  \includegraphics[width=\linewidth]{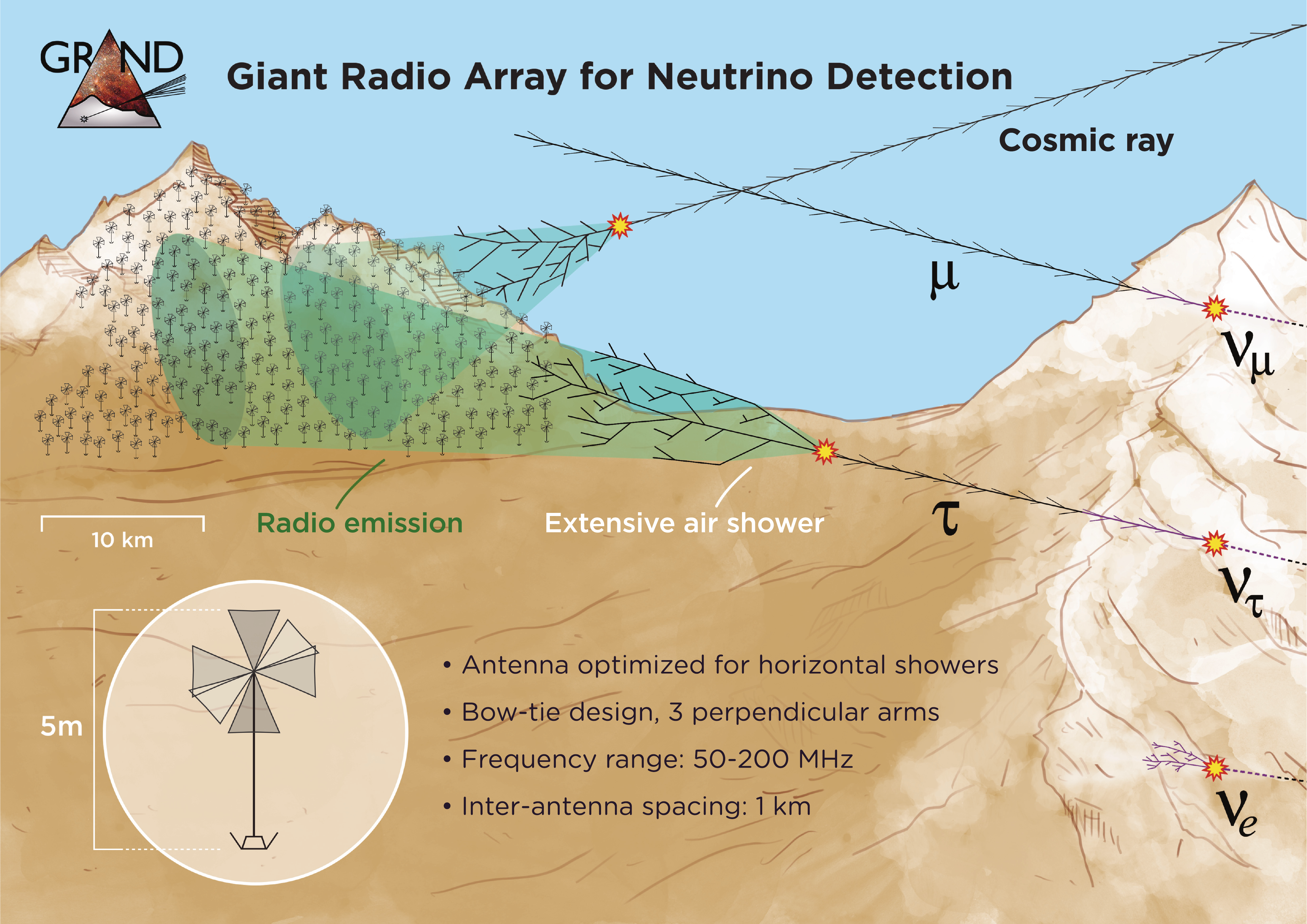}
  \caption{Working principle of the GRAND experiment: sparse radio arrays on gentle slopes will detect air showers.
  GRAND will be the world-largest detector for cosmic rays and search for air showers initiated by neutrino interactions in the Earth. 
  Tau neutrinos interacting in mountains are expected to statistically dominate
  (from \cite{Schroeder_Neutrino2018}).}
  \label{fig_GRAND}
\end{figure*}

\subsection{GRAND}
Also GRAND, the Giant Radio Array for Neutrino Detection, makes use of the large radio footprints of inclined showers \cite{GRAND_ICRC2017}.
GRAND will consist of huge radio arrays searching for ultra-high-energy neutrinos interacting in mountains and subsequently creating air showers (see Fig.~\ref{fig_GRAND}). 
At the same time GRAND serves additional astrophysical and astroparticle science goals \cite{GRAND_Whitepaper2018}. 
In particular, it will detect ultra-high-energy cosmic rays with unprecedented statistics.
In its final stage, $200,000$km$^2$ area will be instrumented with about one antenna station per km$^2$. 
Hence, GRAND will feature an annual exposure for cosmic-ray measurements and photon searches exceeding that of the Pierre Auger Observatory by an order of magnitude.

Building such a large array comes with many technical challenges. 
Thus, a staged design is foreseen with the first prototype of 35 stations currently under construction at the site of the former TREND experiment in China \cite{TREND2011}. 
Subsequently a larger hybrid array of particle detectors and 300 antennas is planned targeting cosmic-ray science goals.
As next step a $10,000\,$km$^2$ radio array will have first discovery potential for neutrinos in the coming decade.
While self-triggering on cosmic rays was already demonstrated, the prototypes need to show that efficiency and purity can be sufficient for the detection of neutrinos. 
Nevertheless, this demonstration has also to be done for the main competitor of GRAND, which is a proposed in-ice radio array for neutrino detection with first prototypes in place (see next section). 
Depending on the results of the Pierre Auger Observatory, already the cosmic-ray science case of GRAND might make it a worthwhile investment, independent of its potential as neutrino detector. 


\subsection{The Square Kilometer Array (SKA)}
The low-frequency array of the Square Kilometer Array in Australia will feature several $10,000$ antennas on less than $1\,$km$^2$. 
Built primarily for radio astronomy it can commensally measure the radio emission of air showers in ultimate detail \cite{Zilles_ARENA2016_SKA}. 
This unprecedented precision may provide a new access to the particle physics of the development of the air showers and, of course, can be used for accurate measurements of Galactic cosmic rays.

The frequency band will be $50-350\,$MHz and, thus, significantly broader than that of previous air-shower arrays. 
Since the dominating Galactic radio noise drops below environmental temperature for frequencies above approximately $150\,$MHz, an antenna with a low noise figure has been developed \cite{SKALAantenna_v2}. 
Using this antenna and the higher frequency band will significantly improve the signal-to-noise ratio and lower the detection threshold compared to LOFAR and other current arrays. 
Moreover, using the SKA antenna might be a good option for future antenna arrays using the same band or a large sub-band.

\begin{figure*}
  \centering
  \includegraphics[width=\linewidth]{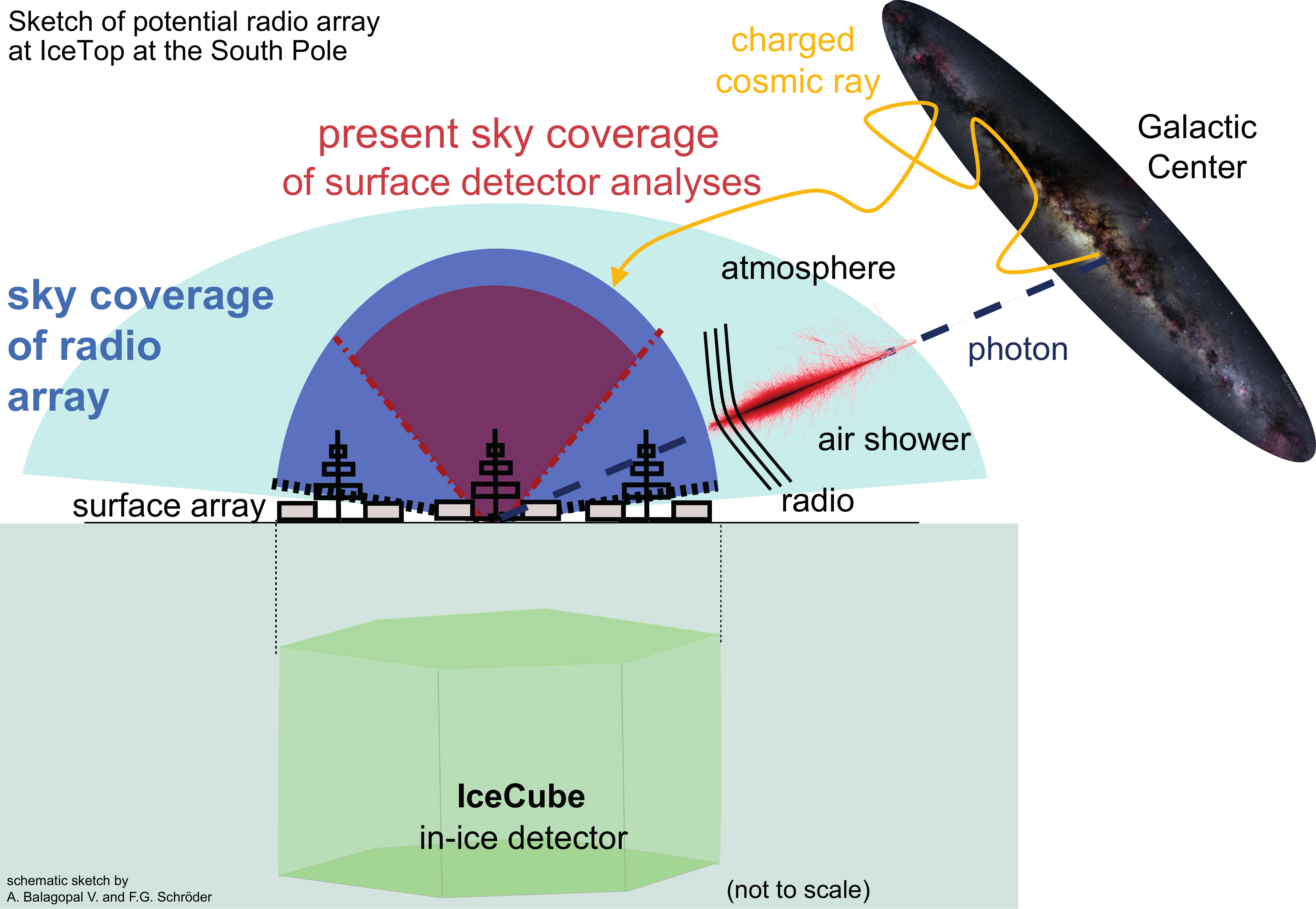}
  \caption{Sky coverage of a potential radio array at the South Pole. 
  Extending existing and planned particle-detector arrays of IceCube will increase the total accuracy for air-shower detection and bring the Galactic Center in the field of view (from \cite{Schroeder_ARENA2018}).}
  \label{fig_RadioAtIceTop}
\end{figure*}

\subsection{Potential Radio Surface Array at IceCube}
IceTop, the air-shower detector of IceCube \cite{IceTopNIM2013}, does not yet feature a radio extension, but previous radio measurements confirmed generally good background conditions at the South Pole (there still is RFI, but much less than in most regions of the Earth) \cite{Boeser_RASTA_ARENA2012}. 
IceTop is the only detector in the southern hemisphere covering the complete energy range of the assumed transition from Galactic to extragalactic cosmic-ray sources. 
Hence, increasing the accuracy of IceTop can be essential to find the most energetic accelerators in our Galaxy. 
In particular, a per-event mass estimator and an increased sky coverage will enable the search for mass-dependent anisotropies. 
Both improvements can be provided by a radio extension (see Fig.~\ref{fig_RadioAtIceTop}).

A CoREAS simulation study on such a potential radio extension revealed that using an optimum detection band of $100-190\,$MHz can drastically lower the threshold down to around $1\,$PeV for inclined showers. 
This will enable a new science goal for radio arrays, namely the search for PeV photons \cite{Balagopal2018}.
The Galactic Center is one of the most promising candidate sources for the most energetic Galactic cosmic rays \cite{HESS_GalacticCenter_Nature2016}, and it is continuously visible at $61^\circ$ zenith angle from the South Pole, i.e., at an inclination ideal for radio detection for air showers. 
If the predicted threshold of $1\,$PeV will be confirmed experimentally, a radio array at the South Pole potentially can measure several PeV photons per year and has discovery potential for the most-energetic source in the Milky Way. 

While the current IceTop array has a relatively high trigger threshold for this zenith angle, an extension by a scintillator array is already planned \cite{IceScint_ICRC2017} - mainly for the calibration of IceTop and as a prototype for a larger veto array.
Depending on its density, this scintillator array can have a threshold of a few PeV or even lower for photons from the Galactic Center and provide a trigger to the radio antennas. 
The gamma-hadron separation would then be done by searching for muon-poor showers with a large radio signal. 
Independent of this science case, the scintillator array provides also a timely opportunity to add radio antennas for marginal additional effort. 
This will boost the total measurement accuracy of IceTop for all arrival directions making it an outstanding detector for Galactic cosmic rays of highest energies \cite{Schroeder_ARENA2018}.

\section{Other media and detector designs}
Apart from surface arrays of antennas, there is another design type of radio detectors for air showers consisting in a single or few stations overlooking large areas. 
This has been realized as balloon payload by ANITA \cite{ANITA_CR_PRL_2010}, and by multi-antenna stations on mountains or on a tower \cite{TAROGE_ICRC2015}. 
While the proof-of-principle of cosmic-ray detection using this design was provided by ANITA a few years ago, it is not yet understood whether this approach can provide sufficient accuracy for cosmic-ray science cases. 
Nevertheless, it might be a suitable option for statistics-driven science cases, in particular the search for EeV neutrinos.

Neutrino search is also the driving science goal for radio detectors using dense media, such as ice. 
Dense media have the disadvantage that the particle showers are so short that geomagnetic radio emission is negligible. 
For the same energy of the primary particle, the radio signal is an order of magnitude weaker in dense media than in air. 
Only for neutrinos as primary particle, this apparent disadvantage is compensated by another effect.
In contrast to cosmic-ray nuclei or photons, the interaction rate of neutrinos is much higher in dense media than in the atmosphere. 
Therefore, when aiming at the same neutrino energies, a radio array in the ice can be much smaller than an antenna array on the ground. 
Hence, instead of simply using dense media (mountains) as neutrino target and detecting air showers, as GRAND is plans to do, the neutrinos can be detected directly in the ice.
This is approach is tested by the ARA and ARIANNA prototype experiments in Antarctica \cite{ARA_PRD_2015, ARIANNA_2016}, and there are plans for a large in-ice radio array at the location of IceCube.

Finally, the lunar regolith provides a huge and suitable target for all type of primary particles \cite{BrayReview2016}, but an experimental proof-of-principle still needs to be provided.

\section{Conclusion}
The current generation of radio arrays has brought the technique to maturity for its application on cosmic-ray science. 
The radio signal calculated by state-of-the-art simulation codes agrees to the measured signals within uncertainties, and simulations are widely and successfully used to interpret measurements. 
Although the radio signal has not yet been understood in ultimate detail, this implies that the  degree of understanding achieved today is sufficient for practically all purposes. 
In particular, it has been demonstrated experimentally that the accuracy of key parameters, such as the cosmic-ray arrival direction, energy, and the position of the shower maximum, is competitive to traditional techniques.
Hereby the radio technique can be used during daytime, clouds, and rain, i.e., the radio technique shares the advantage of a 24/7 duty cycle with the particle-detection technique.

Since the cosmic-ray composition seems to be mixed at all investigated energies \cite{AugerMixedCompAnkle2016, Dembinski_ICRC2017}, the mass-sensitivity of cosmic-ray experiments is much more important than thought a few years, ago. 
Thus, in addition to using radio as possible stand-alone alternative for huge arrays, current particle detector arrays will be enhanced by radio antennas in order to increase their total accuracy. 
For such particle-radio hybrid arrays, the mass sensitivity is not only provided by the $X_\mathrm{max}$ measurement of the radio array, but also by the combination of the radio and the muon signal.

Hence, there are three main use cases for the radio technique that are being realized in the next generation of antenna arrays:
First, a more detailed understanding of the shower physics potentially leading to a further improvement of detection techniques. 
Second, the search for neutrinos which requiring huge areas to be instrumented for reasonable cost. 
This is feasible because the radio footprint covers many km$^2$ for inclined showers. 
Third, the transformation of existing particle-detector arrays into cosmic-ray observatories of unprecedented measurement accuracy.

\section{Acknowledgment}
I would like to thank the LOC of the ISVHECRI for the invitation to this well-organized conference. 
Moreover, I thank David Butler for clarification about the lateral extension of the electrons and positrons contributing to the radio emission.

\bibliography{isvhecri2018}

\end{document}